\documentclass[aps,prl,twocolumn,groupedaddress]{revtex4}
\usepackage{graphics}
\usepackage{amsmath}
\begin{document}

\newcommand*{\CS}{{\cal S}}

\title{Phase sensitive noise in quantum dots under periodic
perturbation}

\author{A. Lamacraft}
\affiliation{Department of Physics, Princeton
University, Princeton, NJ, 08544, USA.}
\date{January 20, 2003}
\email{alamacra@princeton.edu}

\begin{abstract}
We evaluate the ensemble averaged noise in a chaotic quantum dot
subject to DC bias and a periodic perturbation of frequency
$\Omega$. The noise displays cusps at bias $V_n=n\hbar\Omega/e$ that survive the average, even when the period of the perturbation is far shorter than the dwell time in the dot. These
features are sensitive to the phase of the time-dependent scattering
amplitudes of electrons to pass through the system, and thus provide a novel signature of phase-coherent transport that persists into the non-adiabatic limit.
\end{abstract}

\maketitle

Shot noise in a phase coherent conductor is a fundamentally quantum phenomenon. A charge carrier traversing the conductor may leave to each connecting terminal with some probability. Thus noise, like conductance, is in principle sensitive to the quantum dynamics of the charge carriers in mesoscopic systems. As is often the case, however, the shot noise under DC bias may be well described by semiclassical reasoning, with only small quantum corrections for large conductors~\cite{bb}. In this context it of interest to identify situations where the noise has intrinsically quantum mechanical features that are nevertheless significant in a large, but phase coherent conductor. Such a situation is the subject of this Letter.

In 1993 Lesovik and Levitov~\cite{ll} (LL) - building on the work of
Ivanov and Levitov~\cite{il} -  demonstrated that the noise of a
coherent conductor in the presence of a DC bias \textit{and} an AC
external field is a phase sensitive quantity. Specifically, they
showed that the zero temperature noise displays cusps at voltages
$V_n=n\hbar\Omega/e$, multiples of the field frequency $\Omega$,
proportional to a quantity containing the phase of the time-dependent
scattering amplitude of the electrons. For the case of a single
scattering channel, and ignoring the time of flight through the
scatterer, the noise $S$ is a piecewise linear function of bias,
\begin{eqnarray} \label{LLres}
&&\frac{\partial S}{\partial V}=\frac{4e^3}{h}\sum_n
\lambda_n\theta(eV-n\hbar\Omega)\nonumber\\ &&
\lambda_n=\Bigg|\int^{2\pi/\Omega}_0 dt\; t_L^*(t)r_R(t) e^{in\Omega
t}\Bigg|^2\;,
\end{eqnarray}
where $t_L(t)$ and $r_R(t)$ are the amplitudes for transmission from
the left and reflection from the right, respectively. The noise is
related to the uncertainty $\langle\Delta Q_{\tau_0}^2\rangle$ in the
charge
 transported through the system in time $\tau_0$ by
$S=\lim_{\tau_0\to\infty}\langle\Delta
Q_{\tau_0}^2\rangle/\tau_0$. The charge transfer
 statistics of the
of the AC driven system are a mixture of two
 independent Bernoulli
processes, characterized by the integers on
 either side of
$eV/\hbar\Omega$, with attempt frequencies
 $(eV/\hbar-n\Omega)/2\pi$
and  $((n+1)\Omega-eV/\hbar)/2\pi$. This is
 the origin of the
cusps. The key difference from the usual situation
 in shot noise,
corresponding to a purely binomial distribution, is
 that the
inelastic scattering due to the AC field means that the
multiparticle amplitudes for identical fermions come into play. The
fermionic nature of the electrons does not just enter through the
distribution function in the leads.
 
 LL evaluated the $\lambda_n$
for the case of a scatterer in a simply
 connected  loop placed in an
AC magnetic field, neglecting the time of
 flight through the
scattering region. In time-dependent transport, this is usually known as the \emph{adiabatic} limit (note that in the open systems we consider here, there is no trace of level discreteness). The cusps in the noise were
 observed by
Schoelkopf \emph{et al.} in diffusive wires~\cite{skpr}. In this Letter we show that the same phenomenon may be observed
when the dwell time $\tau_{\mathrm{esc}}$ in the scattering region is
not small.  This question arises naturally when the scatterer is a chaotic quantum dot. Here, one may be in the regime $\Omega\gg\tau_{\mathrm{esc}}^{-1}$ while remaining in the `universal' limit $\hbar\Omega\ll E_T$, where $E_T$ is the Thouless energy, something that is not possible in a diffusive wire where $\hbar/\tau_{\mathrm{esc}}\sim E_T$. For micron-sized GaAs quantum dots, with $\hbar/\tau_{\mathrm{esc}}$ in the $\mu eV$ range, a frequency of of 10 GHz, as was applied in the experiment of Ref.~\onlinecite{huibers}, gives $\Omega\tau_{\mathrm{esc}}\sim 100$, so this regime is accessible.

It is not known whether single-particle excitations will be coherent at the high bias required to observe the effect we shall discuss. It should be stressed, however, that the inelastic transitions in the AC field that are responsible are completely coherent. That is, electrons are \emph{coherently excited or de-excited by the applied field}. It is remarkable that this phase-coherent phenomenon survives into the non-adiabatic regime, where weak localization and universal conductance fluctuations do not~\cite{va}.

At low temperatures the principal source
of
 smearing of the cusps will be relaxation through
electron-electron
 scattering. We will neglect this for the remainder
of this paper,
 though this consideration will certainly be relevant
for experiment.

Experimental investigation of time-dependent
transport
 in chaotic dots has so far been concerned with the
construction of
 charge pumps~\cite{switkes}. Noise was measured in a
chaotic cavity
 under DC bias in Ref.~\onlinecite{Oberholzer}. The
measurement of the noise
 in the presence of periodic perturbation
provides a novel probe of
 phase coherent transport. 

\begin{figure} 
\begin{center}
\setlength{\unitlength}{2.8in}
\begin{picture}(1, 0.75)(0,0)
  \put(0,0){\resizebox{1\unitlength}{!}{\includegraphics{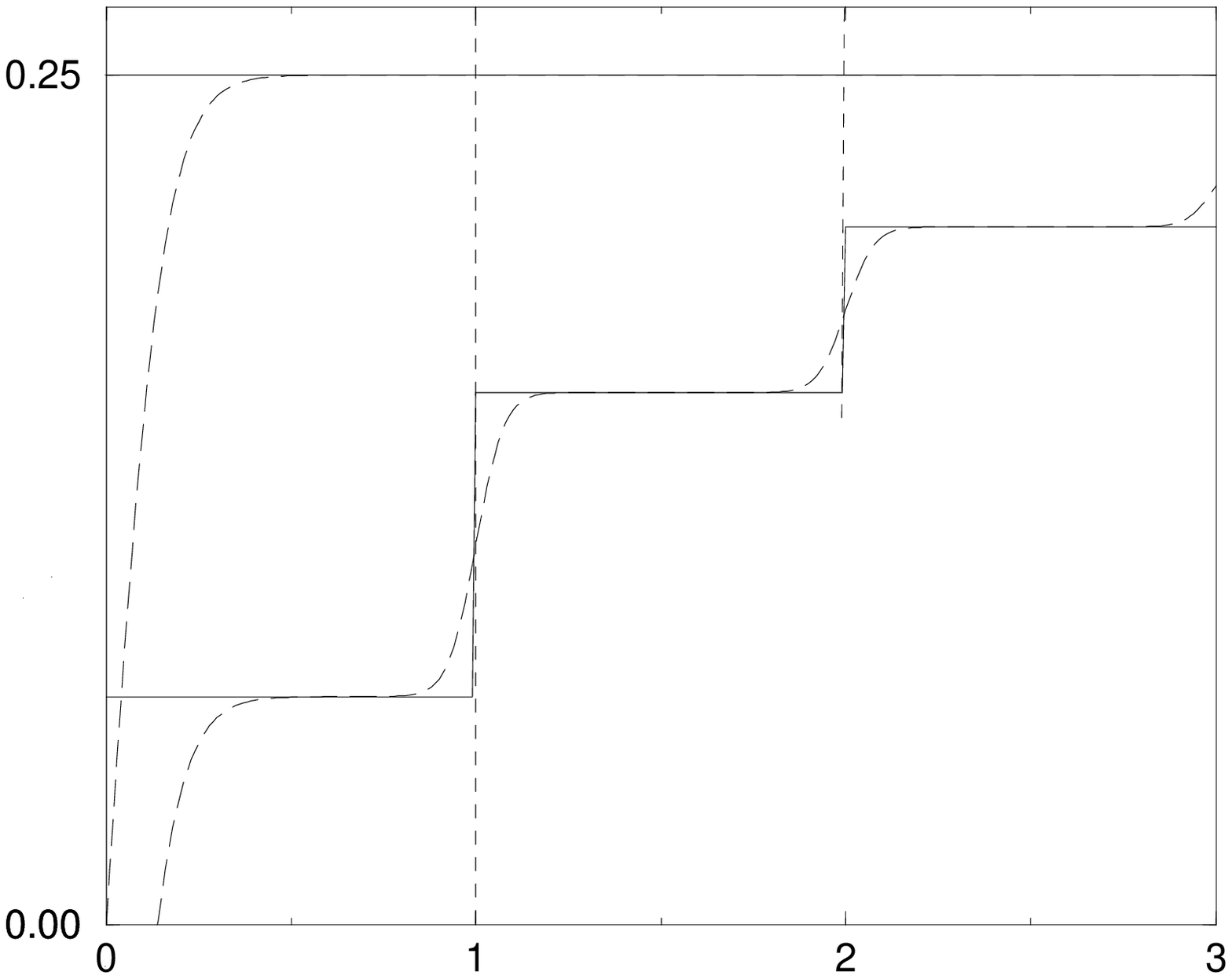}}}
  \put(0.49,0.16){\resizebox{0.4\unitlength}{!}{\includegraphics{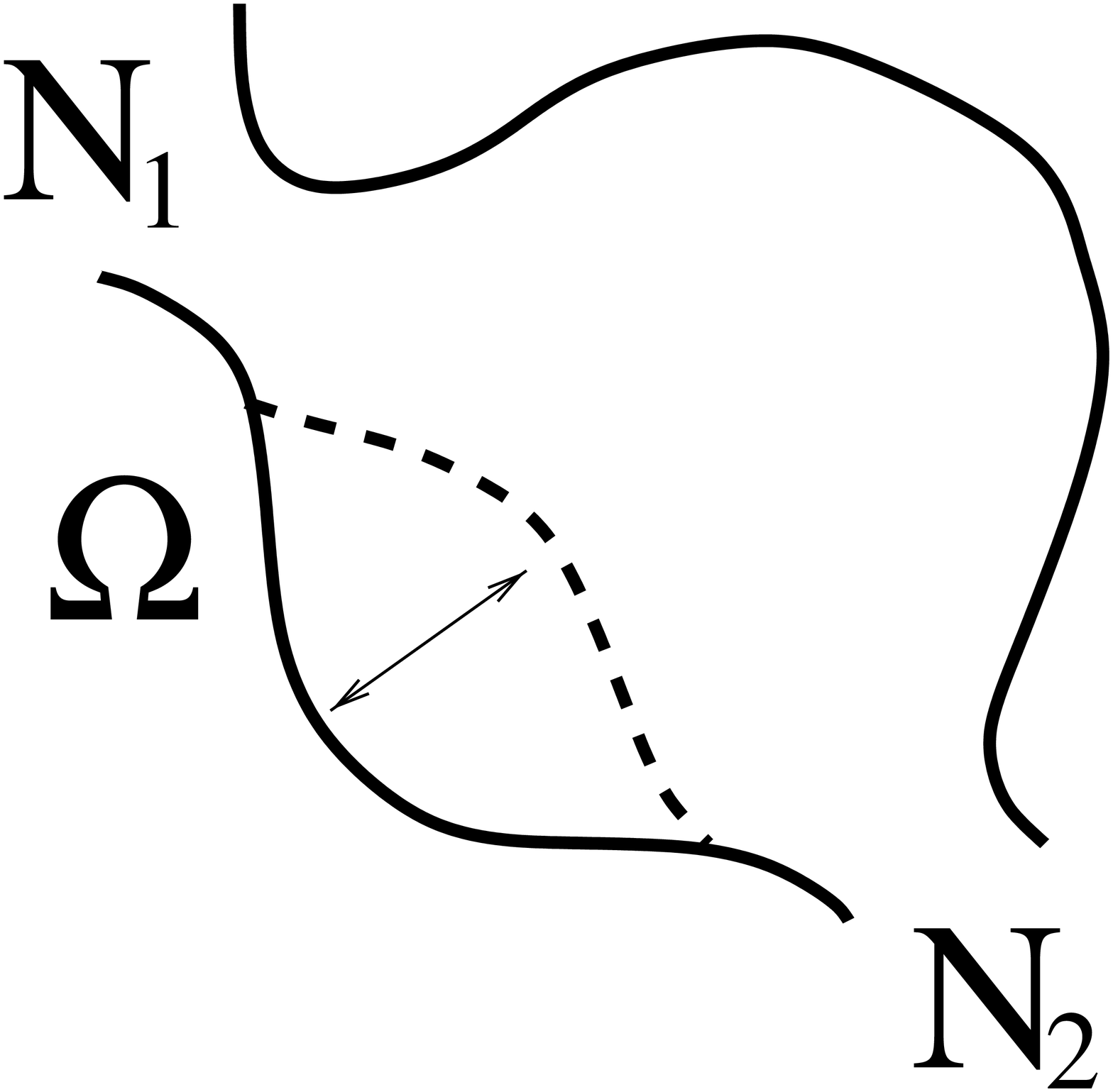}}}
  \put(0.53,0.07){\makebox(0,0)[b]{$eV/\hbar\Omega$}}
  \put(0,0.432){\rotatebox{90}{\makebox(0,0)[t]{$\partial S/\partial
  V\;/e\langle  G\rangle$}}}
\end{picture}
\end{center}
\caption{Derivative of noise with respect to bias  with (lower curves) and without (upper) periodic perturbation for a symmetric ($N_1=N_2$) quantum dot, showing steps at $neV/\hbar\Omega$ at zero temperature (solid line). At finite temperature the steps are smeared ($k_{\mathrm B}T/\hbar\Omega=0.05$, dashed line). We take $C=2$. Inset: experimental realization.\label{fig:cusps}}
\end{figure}

Fig.~\ref{fig:cusps} illustrates our principal finding.  
For a two terminal device with
$N=N_1+N_2$ open channels
 in the presence of an AC field with
$\Omega\gg\tau_{\mathrm{esc}}^{-1}$, the constant slope in the dependence of
the
 noise on bias $S=eIN_1N_2/N^2$ - responsible the famous Fano
factor of
 $1/4$ for $N_1=N_2$~\cite{Jalabert,Oberholzer,nazarov} - is split
to form steps. The familiar
result
 is recovered at large bias. In addition to the
Nyquist noise,
 we find the contribution, depending on `strength' parameter $y$
\begin{eqnarray} \label{res}
S^P&=&\frac{\langle G\rangle\hbar\Omega}{N^2}\{2N_1N_2{\cal
F}(|eV/\hbar\Omega|,y,k_{\mathrm B}T/\hbar\Omega)r\\*\nonumber
&&+(N_1^2+N_2^2){\cal F}(0,y,k_{\mathrm B}T/\hbar\Omega)-N^2k_{\mathrm B}T/\hbar\Omega\}\;.
\end{eqnarray}
${\cal F}(x,y,0)$ is a continuous piecewise linear function of
$x$, with slope 
\begin{eqnarray*}
\frac{\partial{\cal F}}{\partial x}=\textrm{sgn}(n)\left[\frac{1}{2}-y^{|n+1|}\left(\frac{2+y+y^2+\left(1-y^2\right)|n|}{\left(1+y\right)^3}\right)\right]\\*
|n|\leq|x|<|n+1|\;.
\end{eqnarray*}
In Eq.~(\ref{res}) $\langle G\rangle=(2e^2/h)N_1N_2/N$ is the average conductance. The parameter $y\equiv e^{-2\sinh^{-1}\sqrt{1/2C}}$ measures the strength of the perturbation. $C=\tau_{\mathrm{esc}}/\tau_{\mathrm{tr}}$ is the ratio
of the average transition rate induced by the perturbation $\hat V$ in
the Golden Rule approximation ($\tau^{-1}_{\mathrm{tr}}=
2\pi\overline{|V_{\alpha\beta}|^2}/\hbar\Delta$) to the escape rate from
the dot ($\tau_{\mathrm{esc}}^{-1}=N\Delta/2\pi\hbar$), where $\Delta$ is
the level spacing. Such a perturbation may be applied by
periodic deformation of the quantum dot, as illustrated in the inset
to Fig.~\ref{fig:cusps}. The result is derived in the limit $N\gg 1$.
Note that unlike phase coherent signatures
such as weak localization and conductance fluctuations, this effect is
of order unity, not small in the number of channels.

It is possible to understand the size of the steps in $\partial
S/\partial V$  by combining the
results of LL  with the recent description of pumping in dots
in Ref.~\onlinecite{vaa}. The generalization of the result (\ref{LLres}) is
(see Eq.~(\ref{PVB}) below)
\begin{eqnarray} \label{genstep}
\lambda_n=\sum_{m,m'}{\mathrm{tr}}\left\{\Lambda S^{(m)}_{\mu_1}{\cal P}_1
S^{(m')\,\dagger}_{\mu_1}\Lambda S^{(n-m')}_{\mu_1+n\hbar\Omega}{\cal
P}_2 S^{(n-m)\,\dagger}_{\mu_1+n\hbar\Omega}\right\}\;.
\end{eqnarray}
The representation $S^{(m)}_{ij}(E)$ that we use here gives the amplitude to go from channel $j$ to channel $i$ having initial energy $E$ and gaining $m$ quanta to have final energy $E+m\hbar\Omega$. In terms of a general periodic two-time S-matrix $\CS(t,t')=\CS(t+2\pi/\Omega,t'+2\pi/\Omega)$ we have
\begin{subequations} \label{periodic}
\begin{equation}
{\cal S}(t,t')=\sum_m\int\frac{dE}{
2\pi}S^{(m)}_E e^{-iE(t-t')-i\Omega m t}
\end{equation}
\begin{equation}
{\cal S}^{\dagger}(t,t')=\sum_m\int\frac{dE}{
2\pi}S^{(m)\,\dagger}_E e^{-iE(t-t')+i\Omega m t'}\;.
\end{equation}
\end{subequations}
In Eq.~(\ref{genstep}) the matrices ${\cal P}_{1,2}$ project on the
channels in leads 1 and 2 with chemical potentials $\mu_1$ and $\mu_2=\mu_1+n\hbar\Omega$ at the $n^{\textrm{th}}$ step, and the vertex $\Lambda=(N_2{\cal
P}_1-N_1{\cal P}_2)/N$. An ensemble average of Eq.~(\ref{genstep}) in the
limit $\Omega\gg\tau_{\mathrm{esc}}^{-1}$ causes terms with $m\neq m'$ to
vanish, as the S-matrix is only correlated on energy scales of the
order of $\hbar/\tau_{\mathrm{esc}}$ or less. Only the (ensemble averaged)
probabilities $P^{(m)}=N\langle |S^{(m)}_{ij}(E)|^2\rangle$ of an
electron to gain or lose $m$ quanta while passing through the dot survive.
We have
\[\lambda_n=\frac{(N_1N_2)^2}{N^3}\sum_m P^{(m)} P^{(n-m)}\;.\]
We will see that this arises from ${\cal F}(x,y,0)$ of the form
\begin{equation} \label{aveF}
{\cal F}(x,y,0)=\frac{1}{4}\sum_{m.n}\left[|n-x|+|n+x|\right]P^{(m)} P^{(n-m)}\;.
\end{equation}
These formulae have the following meaning. Noise arises from the creation of zero energy particle-hole pairs in the outgoing channels. The quantities $\sum_m P^{(m)} P^{(n-m)}$ are the probabilities for a
pair of energy $n\hbar\Omega$ in the incoming channels to be scattered to a zero energy pair. Their weight in the sum (\ref{aveF}) measures the number of $n\hbar\Omega$ pairs. The disappearance of the noise contribution from $n\hbar\Omega$ pairs as $V$ increases through $n\hbar\Omega/e$ is the origin of the cusps. The statistical independence of these contributions indicates that exchange effects have been lost in the average, as this is the only source of correlation in the noninteracting system.

In the limit of strong pumping $C\gg 1$, where an electron makes many
transitions, its energy diffuses so that $\overline{\Delta
E_t^2}=(\hbar\Omega)^2 t/\tau_{\mathrm{tr}}$. Assuming a Gaussian
distribution of $\Delta E_t$ and combining with the exponential
distribution of dwell times in a quantum dot leads to the estimate of
the  probability
\begin{eqnarray*}
P^{(m)}&\sim& \int^{\infty}_0 dt\; \frac{1}{\tau_{\mathrm{esc}}}\sqrt\frac{\tau_{\mathrm
tr}}{2\pi t}\exp\left(-\frac{t}{\tau_{\mathrm{esc}}}-\frac{m^2\tau_{\mathrm{tr}}}{
2t}\right)\nonumber\\* &\sim&\sqrt\frac{\tau_{\mathrm
tr}}{ 2\tau_{\mathrm{esc}}} \exp\left(-\sqrt{\frac{2\tau_{\mathrm{tr}}}{\tau_{\mathrm{esc}}}}
|m|\right)\;.
\end{eqnarray*}
By substituting into Eq.~(\ref{aveF}), one can recover the given properties of ${\cal F}(x,y,0)$ in the limit $C\gg 1$.

Thus despite the completely randomized
behavior  of the S-matrix on the scale of $\hbar\Omega$, the cusps in
the noise persist provided that the Fermi distribution of the incoming electrons is sufficiently
sharp. What cannot be achieved in this limit is the tuning of the step
size to zero through complete destructive interference of the
multiparticle amplitudes, the direct analogue of the Aharonov Bohm
effect discussed by LL. Now we turn to the details of the
calculation. General formulae for the shot noise in terms of ${\cal S}(t,t')$ have recently been given by Polianski \emph{et
al.}~\cite{pvb}.  They separate the noise into $S=S^N+S^P$, with
$S^N$ the Nyquist noise, and $S^P$ the non-equilibrium contribution.
\begin{widetext}
\begin{eqnarray}   \label{PVB}
S^P= \frac{2e^2}{\tau_0} \int_{0}^{\tau_0} dt dt'\int &&dt_1 dt_2 dt'_1
  dt'_2 \;{\mathrm{tr}}\left[{\cal N}(t_1-t'_2) {\cal
  S}^\dagger(t_1,t)  \Lambda {\cal S}(t,t_2)\tilde{\cal N}(t'_1-t_2)
  {\cal S}^\dagger(t'_1,t')  \Lambda {\cal S}(t',t'_2)
  \right.\nonumber\\*&&\left.  - {\cal
  N}(t_1-t'_2)\delta(t_1-t)\Lambda\delta(t-t_2)\tilde{\cal
  N}(t'_1-t_2)\delta(t_1'-t')\Lambda\delta(t'-t_2')\right]\,.
\end{eqnarray}
\end{widetext}
The matrices ${\cal N}(t)$ and $\tilde{\cal N}(t)$ are given by
\[{\cal N}(t)=\openone-\tilde{\cal N}(t)={\cal P}_1 n_1(t)+{\cal P}_2 n_2(t)\;,\]
where $n_{1,2}(t)$ are the Fourier transforms of the distribution functions in the leads $n_{1,2}(\epsilon)=(e^{(\epsilon-\mu_{1,2})/k_{\mathrm B}T}+1)^{-1}$.  The factor of two in Eq.~(\ref{PVB}) assumes spin degeneracy.

$S^N$ is related to the time-averaged conductance through the fluctuation-dissipation theorem $S^N=2kT\bar G$. The
time-averaged conductance has been analyzed extensively in
Refs.~\onlinecite{va,ykk}. Polianski \emph{et al.} analyzed Eq.~(\ref{PVB}) in
the context of quantum pumps.

The formula (\ref{genstep}) for the steps may be obtained by passing
to the representation (\ref{periodic}). To apply Eq.~(\ref{PVB}) to
chaotic quantum dots, we need to average over realizations. We will
need the two-point correlation function of S-matrices, valid in the
limit $N\gg 1$, in the presence of a time-dependent perturbation
$x(t)\hat V$ (we set $k_{\mathrm B}=\hbar=1$)~\cite{va,pb}
\begin{eqnarray}\label{Save}
\langle{\cal S}_{ij}(\tau,\sigma)  {\cal
S}_{ij}^*(\tau',\sigma')\rangle =&&\delta(\tau-\sigma-\tau'+\sigma')\theta(\tau-\sigma)\nonumber\\* 
&&\times{\cal D}(\tau,\sigma;\tau',\sigma')
\end{eqnarray}
\[{\cal D}(\tau,\sigma;\tau', \sigma') = \frac{\Delta}{
 2\pi}e^{-\tau_{\mathrm{esc}}^{-1}\int_0^{|\tau-\sigma|} d\xi
 \;\left\{1+C\left[x(\sigma+\xi) -
 x(\sigma'+\xi)\right]^2\right\}}\;.\]
Though Eq.~(\ref{PVB}) contains a product of four S-matrices, only the two
point function is required. The non-gaussian connected correlator of
four S-matrices (arising from diagrams containing a Hikami
box)~\cite{vaa,pb} does not contribute. This is because it has the structure $\langle S_{ij}S^*_{kl}S_{mn}S^*_{op}\rangle_{\mathrm{HB}}\propto\delta_{jl}\delta_{km}\delta_{np}\delta_{oi}$, so that its use in Eq.~\ref{PVB} gives factors of $\mathrm{tr}[\Lambda]=0$. Similarly the $\langle \CS\Lambda\CS^\dagger\rangle$ pairings are absent. The choice of current operator
$\hat I=(N_2\hat I_1-N_1\hat I_2)/N$ is responsible for
the traceless $\Lambda$ vertex. Using (\ref{Save}) to average (\ref{PVB}) yields
\begin{widetext}
\begin{eqnarray} \label{avenoise}
S^P=-\frac{2e^2 N_1N_2}{ N\tau_0}&&\int_0^{\tau_0} dt dt' \left(\frac{T}{
2\sinh(\pi T(t-t'+i0))}\right)^2\\*\nonumber
&&\times\left(\left[N_1^2+N_2^2+2N_1N_2\cos(eV(t-t')]\right)\left(
\int_0^\infty {\cal D}(t,t-\xi;t',t'-\xi)d\xi \right)^2-1\right)\;.
\end{eqnarray}
\end{widetext}
We evaluate Eq.~(\ref{avenoise}) for a perturbation $x=\sin\Omega t$ in
the high frequency regime $\Omega\gg \tau_{\mathrm{esc}}^{-1}$. In this limit
the Diffuson may be written
\[{\cal D}(\tau,\sigma;\tau', \sigma') = \frac{\Delta}{
 2\pi}e^{-\tau_{\mathrm{esc}}^{-1}\left(1+
 2C\sin^2\left(\frac{\Omega(\tau-\tau')}{
 2}\right)\right)|\tau-\sigma|}\,,\]
Substituting this into (\ref{avenoise}), and performing the integrals
in the $\tau_0\to\infty$ limit - it is convenient to Fourier transform -  gives the result (\ref{res})
where the function ${\cal F}(x,y,z)$ is
%
%
\[{\cal F}(x,y,z)=\sum_n N(x,z,n) y^{|n|}\left(\frac{1-y}{ 1+y}\right)^2\left(\frac{1+y^2}{1-y^2}+|n|\right)\]
\begin{eqnarray*}
N(x,z,n)=\frac{1}{4}\left[(n-x)\left(\coth\left((n-x)/2z\right)+1\right)\right.\\*
\left.+(n+x)\left(\coth\left((n+x)/2z\right)+1\right)\right]\;.
\end{eqnarray*}
The zero temperature result involves the function ${\cal F}(x,y,0)$
described  earlier. It has the form (\ref{aveF}) with the probabilities given by $P^{(m)}=y^{|m|}/\sqrt{1+2C}$, which coincides with our earlier estimate at $C\gg 1$

We may instead evaluate Eq.~(\ref{avenoise}) in the time domain. In this case, the cusps in the noise arise from poles in the integrand over $t-t'$
 at $(\pm2i\sinh^{-1}\sqrt{1/2C}+2\pi p)/\Omega$
for
 integer $p$, This is the origin of the phase dependence of the
noise:
 $t$ and $t'$ are the initial times in the two electron
trajectories
 that comprise the Diffuson, see Eq.~(\ref{Save}). A
characteristic imaginary time is typical of the quasiclassical
treatment of an AC driven quantum system~\cite{landau}.  By contrast
the zero temperature shot
 noise is determined by a pole at $t-t'=0$, corresponding to
 trajectories which start at the same instant, so that the phase
of the
 time dependent amplitude is irrelevant.

Naturally the cusps also exist in the adiabatic limit considered by LL:
$\Omega\ll\tau_{\mathrm{esc}}^{-1}$. In this case it is convenient to
express the $\lambda_n$ in the form
\[\sum_n \lambda_n \exp\left(-in\Omega\tau\right)=\frac{(N_1N_2)^2}{ N^3}\frac{1+2C\sin^2\left(\Omega \tau/2\right)}{\left(1+4C\sin^2\left(\Omega \tau/2\right)\right)^\frac{3}{ 2}}\;,\]
and one can verify the sum rule $\sum_n \lambda_n=(N_1N_2)^2/N^3$ required to recover the DC result at high bias.

\begin{acknowledgments}
Thanks to Piet Brouwer for discussions. This research is supported in part by the David and Lucille Packard foundation.
\end{acknowledgments}

\end{document}